\documentclass[%
 reprint,
 amsmath,amssymb,
 aps,
 pre,
]{revtex4-1}

\usepackage{color, graphicx}
\usepackage{dcolumn}
\usepackage{bm}
\usepackage{amssymb}
\usepackage{latexsym}
\usepackage{amsfonts}
\usepackage{amsmath}
\usepackage{multirow}
\usepackage{ifthen}

\definecolor{vt}{rgb}{0.9,0,0.6}

\begin{document}

\preprint{APS/123-QED}

\title{Transversal-rotational and zero group velocity modes in tunable magneto-granular phononic crystals}
\author{F. Allein}
\author{V. Tournat}
\author{V. E. Gusev}
\author{G. Theocharis}

\affiliation{LAUM, CNRS UMR 6613, Universit\'e du Maine, Avenue O. Messiaen, 72085 Le Mans, France}

\date{\today}

\begin{abstract}
We report on the design and operation of a 1D magneto-granular phononic crystal composed of a chain of steel spherical beads on top of permanent magnets. The magnetic field of the permanent magnets induces forces in the granular structure. By changing its strength, we can tune the dynamic response of the granular structure. We present experimental results with evidence of coupled transversal-rotational modes, and zero group velocities modes. These observations are well supported by a proposed model taking into account the mechanical coupling between the beads and the magnets by linear stiffnesses and including all degrees of freedom in translations and rotations.
\end{abstract}

\pacs{Valid PACS appear here}
\maketitle



\section{Introduction}

Granular solids are densely packed arrangements of particles that are present in nature or can also be engineered. In the case of a weak cohesion force between particles, the granular solids are referred to as unconsolidated granular media and their elastic properties are strongly driven by the inter-particle contact force laws \cite{hertzbook,Mindlin}. Analogous to other materials having a periodic structure, granular media made of regularly arranged identical particles, such as spheres, constitute phononic crystals, here denoted as granular phononic crystals. Additional to the commonly encountered properties of phononic crystals, e.g. frequency band gaps, negative refraction, an especially appealing property of the granular phononic crystals is their tunable dynamical response. Not only their linear elastic properties are found dependent on the externally applied static force but also their elastic behavior can range from near linear to highly nonlinear, by modifying the ratio of static to dynamic inter-particle displacements \cite{refs}. Consequently, granular phononic crystals have recently played a key role in the study of fundamental wave phenomena, including solitary waves with a highly localized waveform in the case of uncompressed
crystal, discrete breathers and others ~\cite{compacton,gr1,gr2,gr3,gr4,gr5,gr6,chapterG}. They have been also applied in various engineering devices, including among others shock and energy absorbing layers~\cite{dar06,hong05,doney06}, acoustic lenses \cite{Spadoni}, and acoustic rectifiers \cite{Nature11,LAUM}. 

A key point when considering granular phononic crystals is the fact that contact forces between particles are applied at a radius distance from the center of mass, leading in many configurations to the application of moments to the particles, due to non central forces, in addition to central forces. Then, the rotational motion of the particles is excited and the associated degrees of freedom should be accounted for, together with the translational degrees of freedom, in the description of wave propagation and dispersion. This particularity has been already noted and studied in several previously reported experimental and theoretical results. In \cite{NJP}, out-of-plane modes of propagation in granular membranes were theoretically studied and the dispersion of coupled displacement-rotation modes were predicted. 
Localized modes at one end of a one-dimensional granular chain with one rotational and one translational degree of freedom for the particles have been predicted and analyzed in \cite{pichard}, while coupled transversal-rotational modes at the surface of three-dimensional cubic lattices of spheres have been studied in \cite{PRE2016}, showing the existence, among other interesting phenomena, of a zero group velocity point for one of these surface modes of propagation. 
Compared to the variety of these theoretical developments and of the predicted wave phenomena, experimental studies and evidence of the rotational and coupled rotational-translational modes of propagation or even experimental observations of how the rotational degree of freedom influences wave propagation are only very few. 
Transversal-rotational modes of propagation have been identified experimentally in a three-dimensional hexagonal compact arrangement of mm-size spheres in \cite{prlaurelien}. In this case, the propagation along the high symmetry 6-axis of the crystal exhibits a specific dispersion with an isolated frequency band of propagation for rotational-transverse modes. The latter was identified experimentally with a time-frequency analysis of pulsed signals and comparison of outputs from longitudinal and shear sensitive transducers.
Furthermore, the importance of micro-rotations has been recently theoretically and experimentally revealed also in colloidal-based metamaterials \cite{Nick, NickPRB} and in torsional waves in granular chains \cite{PRL_Torsional}.


\section{Motivation and proposed structure}

Ferromagnetic materials can be strongly magnetized in the presence of externally applied magnetic fields. When multiple ferromagnetic particles are in the vicinity of permanent magnetic fields, they interact with each others. Granular chains are line arrangements of particles in contact, which interact by mechanical contact forces. If the particles are made of
ferromagnetic materials, for example stainless steel, then the permanent magnetic field introduces forces between the particles. The forces result in static deformations at the contacts which can be described by contact stiffnesses. By changing the applied magnetic field, the interparticle forces and as a consequence the dynamic response of the granular crystal changes \cite{APL_Allein}.
Based on the above, in this study, we design and operate magneto-granular crystals composed of a chain of steel spherical beads on top of permanent neodymium magnets (NdFeB) in a linear array and in a configuration where the poles of the magnets are alternatively oriented, see Fig. \ref{fig_setup}(a).
The permanent magnets are aligned and glued to an optical table. This proposed structure has the advantage of the readily directed assembling without the need of external static load.

\subsection{Linearized equations of motion}

\begin{figure}[ht!]
\centering
\includegraphics{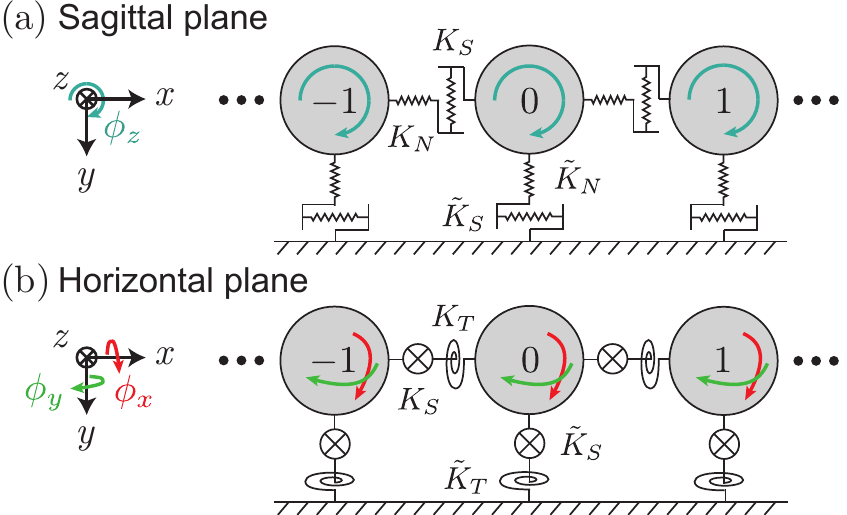}
\caption{\label{fig_mass_spring_xy_xz_3D} (a) Representation of the infinite granular chain coupled with a rigid substrate.  Displacement and rotation motions are separated in (a) the sagittal plane (displacements along $x$ and $y$, rotation relative to $z$-axis) and (b) the horizontal plane (rotations relative to $x$ and $y$-axes, displacement along $z$).}
\normalsize
\end{figure}

To describe the three-dimensional dynamics of a linear granular chain and its mechanical coupling with a rigid substrate, we consider the model as shown in Fig.\ref{fig_mass_spring_xy_xz_3D}.
The possible motions are presented into two decoupled planes, a "sagittal" plane ($x$,$y$) and a "horizontal" plane ($x$,$z$).
The model considers 6 degrees of freedom for each $n$-th sphere  of the chain; 3 displacements along $j=x,y,z$-direction (${u}_{n,j}$) and 3 rotations (${\phi}_{n,j}$). 
Between two adjacent spheres, we consider normal, shear and torsional coupling characterized by constant rigidities $K_N$, $K_S$ and $K_T$ respectively. 
The coupling of the linear granular chain with the substrate is characterized by the normal, shear and torsional constant rigidities $\tilde{K}_N$, $\tilde{K}_S$, $\tilde{K}_T$, respectively.
Note that in this study we ignore a possible bending coupling between the particles verified to be very weak \cite{ultasonics_Zheng}.

Assuming small displacements and rotations
the equations of motion of the zeroth particle are given by:

\begin{subequations}
\begin{eqnarray}
m \ddot{u}_{0,x} & = & K_N \left( u_{1,x} - 2 u_{0,x}  + u_{-1,x}  \right) \\
&&  - \tilde{K}_S \left( u_{0,x} - R\phi_{0,z} \right),  \label{subeq_explmx}
\nonumber \\
m \ddot{u}_{0,y} &  = & K_S \left( u_{1,y} - 2 u_{0,y}  + u_{-1,y} \right) - \tilde{K}_N u_{0,y}  \\
&&  + K_S R \left( \phi_{-1,z} - \phi_{1,z} \right),  \label{subeq_explmy}
 \nonumber \\
m \ddot{u}_{0,z} &  = & K_S \left( u_{1,z} - 2 u_{0,z}  + u_{-1,z} \right) \\
&& - \tilde{K}_S \left( u_{0,z} - R\phi_{0,x} \right)  
  + K_S R \left( \phi_{-1,y} - \phi_{1,y} \right),  \label{subeq_explmz}
 \nonumber \\
 I \ddot{\phi}_{0,x} &  = & K_T \left( \phi_{1,x} - 2 \phi_{0,x}  + \phi_{-1,x}  \right) \\
 &&+ \tilde{K}_S R \left( u_{0,z} - R\phi_{0,x} \right),\label{subeq_explpsix} 
 \nonumber \\
 I \ddot{\phi}_{0,y} &  = & -K_S R^2 \left( \phi_{1,y} + 2\phi_{0,y} + \phi_{-1,y} \right) \\
&&  + K_S R \left(u_{1,z} - u_{-1,z} \right) - \tilde{K}_T\phi_{0,y} , \nonumber\label{subeq_explpsiy}
\nonumber \\
I \ddot{\phi}_{0,z} &  = & -K_S R^2 \left( \phi_{1,z} + 2\phi_{0,z} + \phi_{-1,z} \right) \\
&&  + K_S R \left(u_{1,y} - u_{-1,y} \right) + \tilde{K}_S  R \left( u_{0,x} - R\phi_{0,z} \right), \nonumber\label{subeq_explphiz}
\end{eqnarray}
\label{eq_setmotion}
\end{subequations}
where $m$ is the mass, $R$ is the radius and $I$ is the momentum of inertia of the particle (for the particular case of homogeneous spheres $I = \frac{2}{5}mR^2$). 

\subsection{Contact rigidities\label{section_rigidity}}

For the case of macroscopic granular chains made of elastic spheres, one obtains the normal, shear and torsional rigidities by Hertzian contact mechanics \cite{hertzbook, Mindlin}.
In particular, for the case of two identical spheres, we obtain:  
\begin{eqnarray}
K_N  & = & \left( \dfrac{3R}{4}F_0 \right)^{1/3}{E_1}^{2/3}(1-{\nu_1}^2)^{-2/3},\\
K_S & = &  (6F_0 R)^{1/3} {E_1}^{2/3} \dfrac{(1-{\nu_1}^2)^{1/3}}{(2-\nu_1)(1+\nu_1)}, \\
K_T  & = &  \left( 2 R \right) \left( 1 - \nu_1\right) F_0.
  \end{eqnarray}
$F_0$ denotes the static load along the line that passes through the centers of the spheres (in our case along the $x$-axis).  
The normal, shear and torsional rigidities between each sphere and magnet considering two different materials can be found in \cite{hertzbook, popovbook} and are given by
\begin{eqnarray}
\tilde{K}_N & = &\dfrac{3}{2} \left[ \dfrac{4 E_1 E_2 \sqrt{R}}{3E_2(1-{\nu_1}^2)+3E_1(1-{\nu_2}^2) }\right]^{2/3}  {F}^{1/3},\\
\tilde{K}_S & = &\left(6F R\right)^{1/3} \left[\dfrac{E_2(1-{\nu_1}^2)+E_1(1-{\nu_2}^2)}{E_1E_2}\right]^{1/3}\nonumber \\
&  &\times \dfrac{2E_1E_2}{E_2(2-\nu_1)(1+\nu_1)+E_1(2-\nu_2)(1+\nu_2)},\\
\tilde{K}_T & = & (2 F R)  \dfrac{E_2(1-\nu_1^2) + E_1(1- \nu_2^2)}{E_1+E_2+E_2\nu_1+E_1\nu_2},
  \end{eqnarray}
where indices 1 and 2 refer to the bead and magnet respectively; $E$ is the Young's modulus and $\nu$ the Poisson ratio. $F$ denotes the static load along the axis formed by the center of the bead and the center of the magnet. 
The forces $F_0$ and $F$ induced by the external magnetic field can be estimated by measuring
the pulling force required to separate two adjacent particles using a dynamometer. Experimentally, we measured the two characteristic normal forces to be $F_0=4 \pm 0.15$~N and $F=10 \pm 0.15$~N for cylindrical permanent magnets of remanent magnetization $B_r = 1.37$~T
at contact with the spheres.
For these estimated forces, the values of the stiffnesses are $K_N = 1.05\cdot10^{7}$ N/m, $K_S = 8.62\cdot10^{6}$ N/m, $K_T = 4.44\cdot 10^{-2}$~m$\cdot$N,
 $\tilde{K}_N = 1.63\cdot10^{7}$ N/m, $\tilde{K}_S = 1.38\cdot10^{7}$ N/m and $\tilde{K}_T = 1.16\cdot 10^{-1}$ m$\cdot$N.

\subsection{Dispersion Relation}

The set of Eqs.~(\ref{eq_setmotion}) can be split into two decoupled sets corresponding to the particle displacements in two orthogonal planes, sagittal and horizontal. The motion in the "sagittal" plane ($x$,$y$) includes two displacements along $x$ and $y$ directions while rotation is around $z$-axis. The motion in the "horizontal" plane ($x$,$z$) is formed of one displacement along $z$ direction and two rotations around $x$ and $y$-axis. Solutions of the set of Eqs.~(\ref{eq_setmotion}) are plane waves propagating in the $x$-direction for each plane, the indices "S" and "H" respectively referring to the sagittal and to the horizontal plane:
\begin{subequations}
 \begin{eqnarray}
 {\bf V^s_{n}}  & = & \left(
\begin{array}{c}
u_{n,x}(x,t) \\
u_{n,y}(x ,t) \\
\varPhi_{n,z}(x ,t)
\end{array}\right)
=  {\bf v^s} e^{i\omega t - i k x_n},\\
 {\bf V_n^{\textbf{\tiny{H}}}}  & = &  \left(
\begin{array}{c}
u_{n,z}(x ,t) \\
\varPhi_{n,x}(x,t) \\
\varPhi_{n,y}(x ,t)\\
\end{array}\right)
 =  {\bf v^{\textbf{\tiny{H}}}} e^{i\omega t - i k x_n},
  \end{eqnarray}
   \label{eq_Vn}
\end{subequations}
considering the new variables $\varPhi_{x}~=~R \phi_{x}$,  $\varPhi_{y}~=~R\phi_{y}$ and $\varPhi_{z}~=~R\phi_{z}$.
Here, $k$ is the complex wave number in the $x$-direction,
${\bf v}$ the amplitude vector and $\omega$ is the angular frequency. For each plane, Eqs.~(\ref{eq_Vn}) are developed around the equilibrium position $x_0$ of the central particle ${\bf V_n} = {\bf v} e^{i\omega t - i k x_0}e^{- i k \Delta x_n}$, where $\Delta x_n = x_n-x_0$ is the relative coordinate between the central particle and the $n$-th particle. Substituting Eqs.~(\ref{eq_Vn}) into the set of Eqs.~(\ref{eq_setmotion}) leads to two eigenvalue problems
\begin{eqnarray}
 {\bf D^s} {\bf v^s} = -\Omega^2 {\bf v^s}, \label{eq_eigen1}\\
  {\bf D^{\textbf{\tiny{H}}}} {\bf v^{\textbf{\tiny{H}}}} = -\Omega^2 {\bf v^{\textbf{\tiny{H}}}}, 
 \label{eq_eigen2}
  \end{eqnarray}
where $\Omega = \omega / \omega_0$ is the reduced frequency with $\omega_0~=~\sqrt{K_S/ m}$.
The elements of the $3\times3$ dynamical matrix ${\bf D^s}$ for the sagittal plane and for the horizontal plane ${\bf D^{\textbf{\tiny{H}}}}$ are
 \begin{widetext}
 \begin{eqnarray}
 {\bf D^s} = \left(\begin{array}{ccc}-4 \eta_1 \sin^2(q) - \eta_2 & 0 & \eta_2 \\0 & -4 \sin^2(q) - \eta_3 & 4 \textsl{i} \sin(q) \cos(q) \\\eta_2 \textsl{p} & -4 \textsl{i} \textsl{p} \sin(q) \cos(q) & -4 \textsl{p} \cos^2(q) - \eta_2 \textsl{p}\end{array}\right)
  \end{eqnarray}
   \begin{eqnarray}
   {\bf D^{\textbf{\tiny{H}}}}  = \left(\begin{array}{ccc}- 4 \sin^2(q) - \eta_2 & \eta_2 & 4 \textsl{i} \sin(q) \cos(q) \\\eta_2 \textsl{p} & -4 \eta_4 \textsl{p} \sin^2(q)- \eta_2 \textsl{p} & 0 \\-4 \textsl{i} \textsl{p} \sin(q) \cos(q) & 0 &  -4 \textsl{p} \cos^2(q)- \eta_5 \textsl{p} \end{array}\right)
     \end{eqnarray}
 \end{widetext}
where $\eta_1 = \dfrac{K_N}{K_S}$, $\eta_2 = \dfrac{\tilde{K}_S}{K_S}$, $\eta_3 = \dfrac{\tilde{K}_N}{K_S}$, $\eta_4 = \dfrac{K_T}{K_SR^2}$, $\eta_5 =~\dfrac{\tilde{K}_T}{K_SR^2}$, $\textsl{p} = \dfrac{mR^2}{I}$ and $q = \dfrac{k a}{2}$, where $a = 2R$ corresponds to the lattice constant. The ratio $\textsl{p}$ characterizes the mass distribution in the spherical particle. In our model we consider homogeneous particle ($\textsl{p}$ = 2.5) but by modifying this ratio the dispersion curves could be strongly modified.

\begin{figure*}[ht!]
\centering
\includegraphics{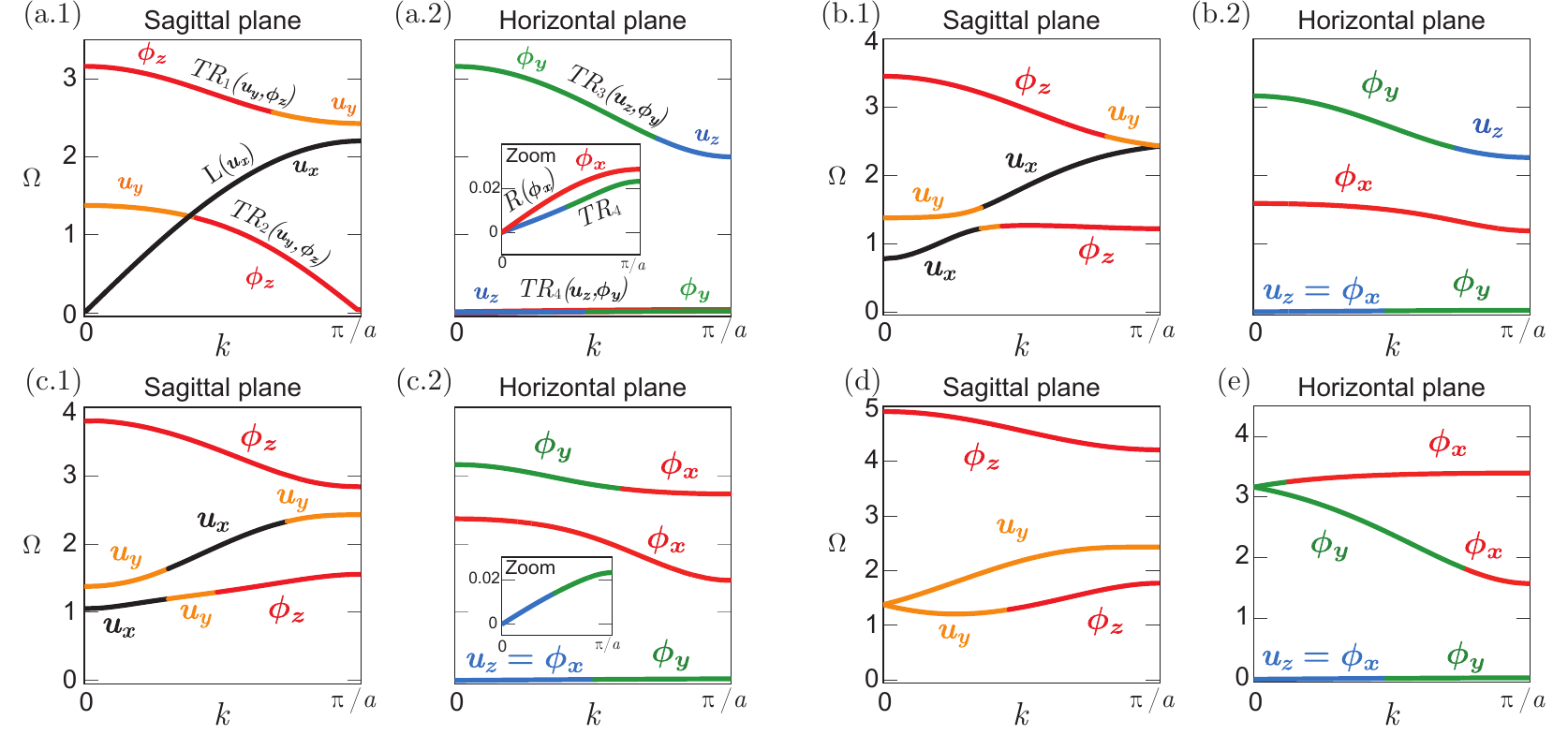}
\caption{\label{fig_dispersion_simu} Dispersion curves of a infinite granular chain coupled to a rigid substrate, as a function of the free parameter $\eta_2 =  \tilde{K}_S/K_S$. 
(a) $\eta_2 =  0$, no shear coupling, (b) $\eta_2 = 0.72$ (c) $\eta_2 = 1.60$  corresponds to the Hertz prediction with the force created by the magnet of $B_r$ (1.37~T) remanent magnetization, (d) $\eta_2 = 4.56$ (e) $\eta_2 = 2.86$. The values of the other parameters are $\eta_1= 1.21$, $\eta_3= 1.89$, $\eta_4= 8.18\cdot 10^{-5}$ and $\eta_5=2.14\cdot 10^{-4}$. $\Omega = \omega/ \omega_0$ is the reduced frequency with $\omega_0 = \sqrt{K_S/m}$.}
\normalsize
\end{figure*}

The solution of the eigenvalue problem given by Eqs~(\ref{eq_eigen1})-(\ref{eq_eigen2}) gives the $\Omega-k $ dispersion relation. 
In Fig.~\ref{fig_dispersion_simu} we show some cases of the dispersion relation of the structure within the first Brillouin zone for waves propagating to the right ($\Re(k) > 0$) varying the $\eta_2$ ratio, which corresponds to modifying the shear rigidity $\tilde{K}_S$ between bead and substrate. Our choice to consider the $\tilde{K}_S$ as a free parameter is discussed in the experimental section. The values of the other parameters are $\eta_1= 1.21$, $\eta_3= 1.89$, $\eta_4= 8.18\cdot 10^{-5}$ and $\eta_5=2.14\cdot 10^{-4}$. These correspond to experimental relevant values taking into account the measured forces and the Hertzian theory (see section~\ref{section_rigidity}). 
In the panels presented in columns 1 and 3, we plot the dispersion relation of sagittally polarized waves that include longitudinal motion ($u_x$), transversal motion along $y$-direction ($u_y$) and  rotational motion around $z$-axis ($\phi_z$), while in panels in columns 2 and 4, we display the dispersion relation of horizontal polarized waves that include transversal motion along $z$-direction ($u_z$)
and rotational motions around x-axis ($\phi_x$) and y-axis ($\phi_y$). The predominance of each motion is labeled in Fig.~\ref{fig_dispersion_simu}.  

The analytical values of the frequencies of the modes at $k=0$ and $k = \pi/a$ are explicitly given for the modes polarized in the sagittal plane ($u_{x}$, $u_{y}$, $\phi_z$) as follows:
For $k = 0$, $\Omega^2_2= \eta_3$, $\Omega^2_{1,3} = \frac{1}{2}\left[4\eta_1+\eta_2(1+\textsl{p}) \pm\sqrt{(4\eta_1+\eta_2(1+\textsl{p}))^2-16\eta_1\eta_2\textsl{p}} \right]$, while for $k = \pi/a$, $ \Omega^2_1= 4+\eta_3$, $\Omega^2_{2,3}= \frac{1}{2}\left[4\eta_1+\eta_2(1+\textsl{p}) \pm\sqrt{(4\eta_1+\eta_2(1+\textsl{p}))^2-16\eta_1\eta_2\textsl{p}} \right]$.
For the modes polarized in the horizontal plane ($u_{z}$, $\phi_{x}$, $\phi_{y}$), one has
for $k = 0$, $\Omega^2_1=0$, $\Omega^2_{2}= \eta_2(1+\textsl{p})$, $\Omega^2_{3}= \textsl{p} \Big(4+\eta_5 \Big)$ while for $k = \pi/a$, $\Omega^2_1= \eta_5 \textsl{p}$, $\Omega^2_{2,3}=\frac{1}{2} \Big( 4\eta_4\textsl{p} + \eta_2(1+\textsl{p})+4\pm \Big[(4\eta_4\textsl{p} + \eta_2(1+\textsl{p})+4)^2-16 \textsl{p}(4\eta_4+\eta_2 +\eta_2 \eta_4) \Big]^{1/2} \Big)$.

In Fig.~\ref{fig_dispersion_simu}(a.1-2) we consider the case in which the granular chain is coupled to a substrate only through the $\tilde{K}_N$ and $\tilde{K}_T$, namely $\tilde{K}_S=0$. 
For the sagittal plane motion, the longitudinal branch $L(u_x)$, with displacement only along the $x$-axis, is uncoupled to the two other transversal-rotational branches denoted  by $TR_{1,2}(u_y, \phi_z)$, with displacement along $y$-axis and rotation around the $z$-axis. Similarly, a pure rotation branch $R(\phi_x)$ is uncoupled to the two $TR_{3,4}(u_z,\phi_y)$. We have to notice that the $TR_{1,2}$ and $TR_{3,4}$ are not identical between the sagittal ($x$,$y$) and horizontal ($x$,$z$) plane.

When the shear coupling between the chain and the rigid substrate is activated, each branch in the dispersion diagram contains all the three components of the relevant mode family.
In Fig.~\ref{fig_dispersion_simu}(b.1-2) we consider the case of $\eta_2 = 0.72$. According to the analytical expressions for the frequencies of the modes at $k=\pi/a$  in the sagittal plane,
this corresponds to the case at which two branches degenerate at the edge of the Brillouin zone. For this case, we also observe that the first two branches in the sagittal plane show an anti-crossing effect which leads to the existence of a zero group velocity (ZGV) mode at a finite wavelength, $k\approx \pi/(2a)$.  
In particular, increasing the parameter $\eta_2$ from zero to a finite value, we observe that the dominant component of the first branch, at the crossing point with the second branch, is not the longitudinal $u_x$ any more but the transversal $u_y$. This explains the avoiding crossing of these two branches and the existence of a ZGV mode.  One can also observe the existence of a band gap from zero up to the value $\Omega_{1}$. Since this value depends on the free parameter $\eta_2$, it can be tuned by  the shear coupling between the granular chain and the substrate.
In Fig.~\ref{fig_dispersion_simu}(c.1-2), we present the case using the values of the stiffnesses given in section~\ref{section_rigidity}, thus $\eta_2=1.60$. The three branches in each plane are smooth and they do not exhibit ZGV. 
In Fig.~\ref{fig_dispersion_simu}(d) and Fig.~\ref{fig_dispersion_simu}(e), we consider the cases of $\eta_2=4.56$ (sagittal plane) and $\eta_2=2.86$ (horizontal plane) respectively.  
Accidental degeneracies at $k=0$ are present in both cases as well as a ZGV mode at $k\approx 0.26 \pi/a$  (see Fig.~\ref{fig_dispersion_simu}(d)). To conclude this section, we emphasize that the granular chain coupled to a rigid substrate results in rich dispersion behavior including ZGV modes at finite wavelength and accidental degeneracy. 
Dispersion engineering by tuning some of the contact stiffnesses can thus lead to interesting wave phenomena. In the following section, we present the proposed experimental setup and the observed dispersion relations.

\section{Experimental Results and Discussion}

\subsection{Linear spectrum}

\begin{figure}[ht!]
\centering
\includegraphics[width= 0.45\textwidth]{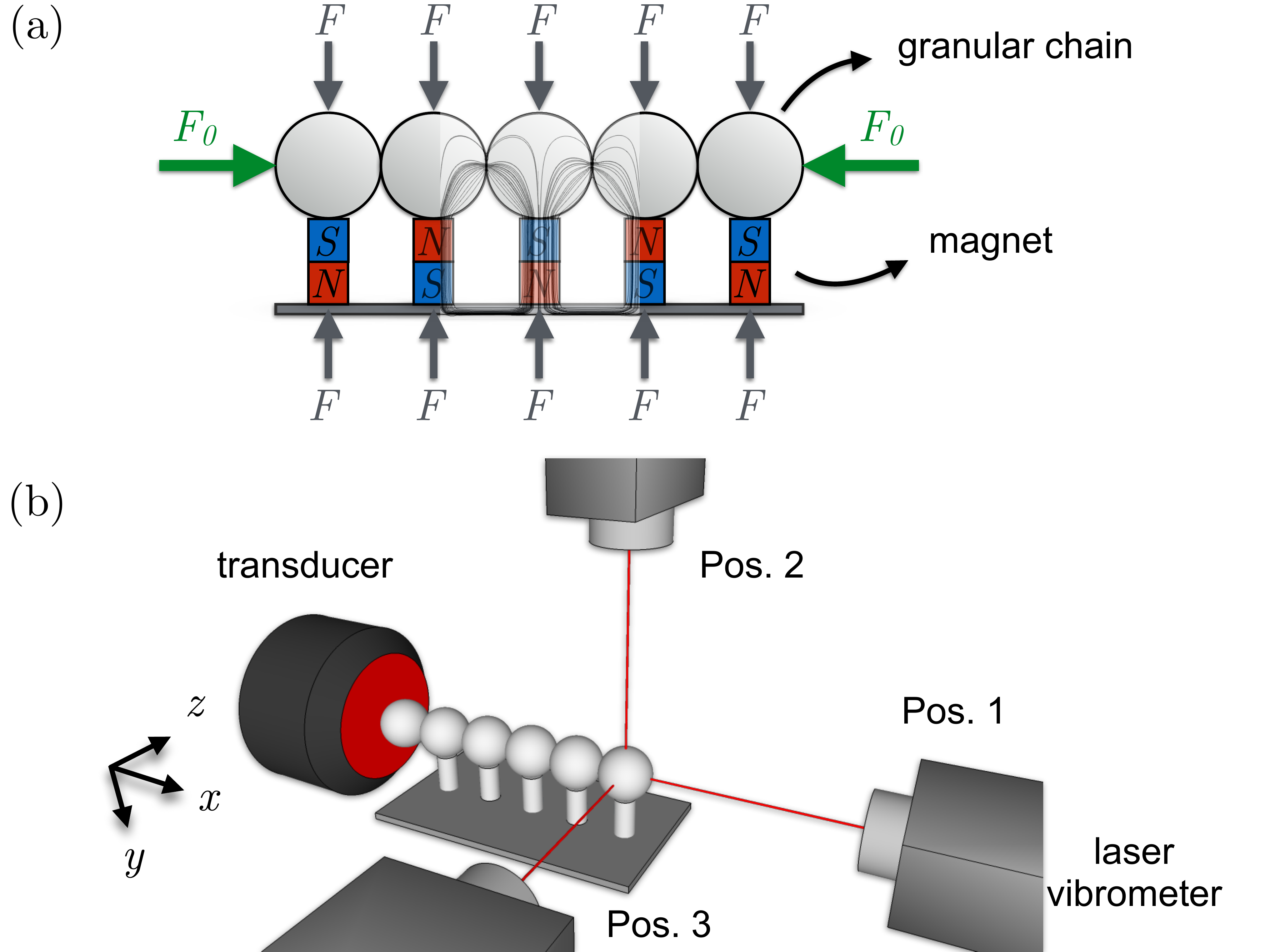}
\caption{\label{fig_setup} (a) Schematics of the granular chain on top of permanent magnets. Forces and current lines induced by the magnetic field are superposed. (b) Schematics of the experimental setup. The configuration allows us to drive the chain along $x$, $y$ or $z$ directions. 
We place the vibrometer in 3 different positions, Pos. 1, 2 and 3 to detect displacement along the $x$, $y$ and $z$ axis, respectively.}
\normalsize
\end{figure}

The experimental setup is composed of a chain of stainless steel beads of 15.875~mm diameter, in direct contact to permanent NdFeB magnets in a linear array configuration. 
The material characteristics of the steel 440C beads are the following: density $7650 kg/m^3$, Young's modulus $E_1=200$ GPa, Poisson ratio $\nu_1=0.3$.
For the magnets, we have density $7600 kg/m^3$, Young's modulus $E_2=160$ GPa, Poisson ratio $\nu_2=0.24$ and remanent magnetization $B_r = 1.37$~T.
A schematics of the setup configuration is shown in Fig.~\ref{fig_setup}(a). 
The chain is driven at one extremity in three independent directions ($x$,$y$,$z$) by using two kinds of transducers (a longitudinal \emph{Panametrics V3052} and a shear \emph{Panametrics V1548}).
A bead is glued to the transducers to ensure the same contact between the driver and the chain and between beads. The excitation signal is a sweep sine of $65$~s duration with a span from $100$~Hz to  $15$~kHz. The measurement of a particle velocity is made by a laser vibrometer \emph{Polytec OFV-503} with a sensibility of $5$~mm/s/V and an averaging is performed during the acquisition. We place the laser vibrometer in three different positions to be able to detect all directions of displacement of the bead. The experimental setup configurations are presented in Fig.~\ref{fig_setup}(b). 

\begin{figure}[ht!]
\centering
\includegraphics{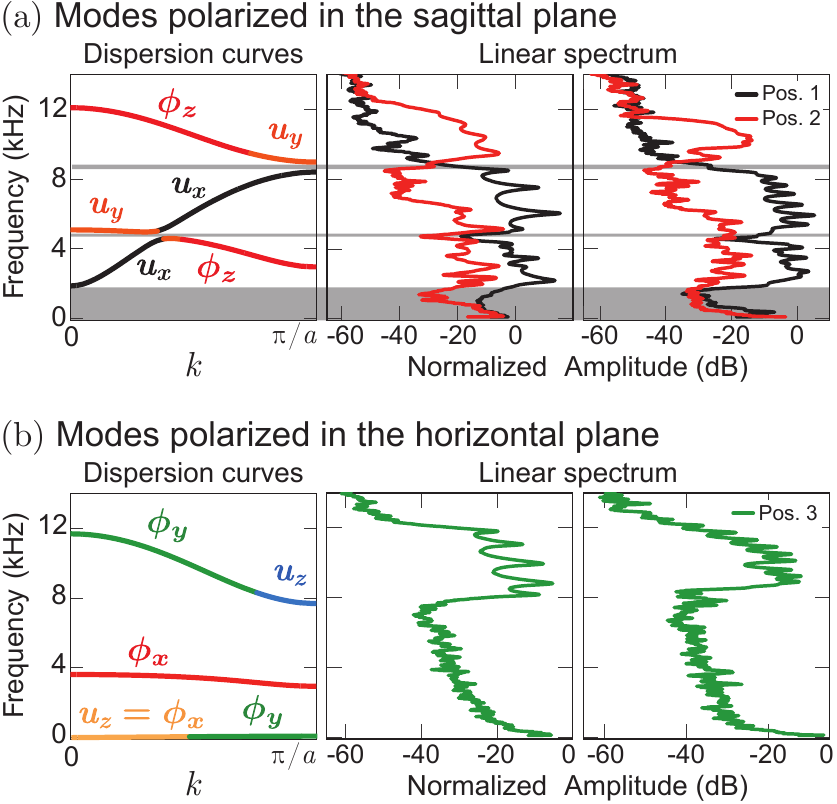}
\caption{\label{fig_Dis_exp_B2_Magnet_15beads} Dispersion curves (left) for the infinite granular chain considering the experimentally measured and estimated forces $F_0 = 4$~N ($K_N$,$K_S$,$K_T$), $F = 10$~N ($\tilde{K}_N$,$\tilde{K}_T$) and experimentally fitted $\tilde{K}_S$. Linear spectrum of the signal (velocity amplitude) detected by the laser vibrometer for a chain composed of 5 beads (center) and 15 beads (right) (a) in the sagittal plane ($x$,$y$) and (b) in the horizontal plane ($x$,$z$). 
The measurement is made on the last bead of the chain. We measure in the same direction than the polarization of the excitation. Pos. 1 (black line) measurement along $x$-axis, Pos. 2 (red line) along $y$-axis, and Pos. 3 (green line) along $z$-axis. The grey zones correspond to the theoretical forbidden propagation band-gaps.}
\normalsize
\end{figure}

The panels in the center and right  of Fig.~\ref{fig_Dis_exp_B2_Magnet_15beads} present the experimental results of a chain with 5 and 15 beads respectively, on top of the cylindrical permanent magnets. From the dispersion curve presented in Fig.~\ref{fig_dispersion_simu}(c.1-2), we expect a large forbidden band of propagation from 0 up to $3880$~Hz  
  for the longitudinal motion ($u_x$). The width of this forbidden gap is governed mostly by the shear rigidity between each bead and the substrate. 
The Hertzian prediction of the shear rigidity considering a normal load of $F = 10$~N gives $\tilde{K}_S = 1.38 \cdot 10^7$~N/m.
  As we can see on the middle panel of Fig.~\ref{fig_Dis_exp_B2_Magnet_15beads}(a), (linear spectrum measured in the Pos.~1 which is sensitive to $u_x$ motion) the experimentally measured band gap is up to $1870$~Hz. Considering the shear stiffness between the granular chain and the substrate as a free parameter, this band gap is produced by
using $\tilde{K}_S=2.36 \cdot 10^6$~N/m, namely an order of magnitude smaller. 
A similar discrepancy of the shear stiffness (Hertzian expected and experimentally measured) has been also observed in \cite{microsilica}, for the case of micro-sized silica particles.
Using this fitted parameter for $\tilde{K}_S$, a very good agreement between experimental results and the predicted allowed and forbidden bands of propagation is achieved.
In particular, three band gaps (denoted by the grey area in Fig.~\ref{fig_Dis_exp_B2_Magnet_15beads}(a)) have been observed. 
At $f=4570$ Hz, a dip in the linear spectrum is noticed (a narrow band gap) to be in agreement with the frequency gap around the theoretical ZGV mode for $k\approx 0.38 \pi/a$.
This is an experimental evidence of the existence of the ZGV mode. 
For the modes polarized in the horizontal plane, the lowest branch can not be excited by our transducer so we can not measure it. The middle branch of the dispersion relation curve has a very strong predominance of rotation ($\phi_x$) and only a small translation component. Thus, it can not be detected by the laser vibrometer which permits to detect motion of the reflecting surface along the optical beam direction.

\begin{figure}[ht!]
\centering
\includegraphics[width= 0.4\textwidth]{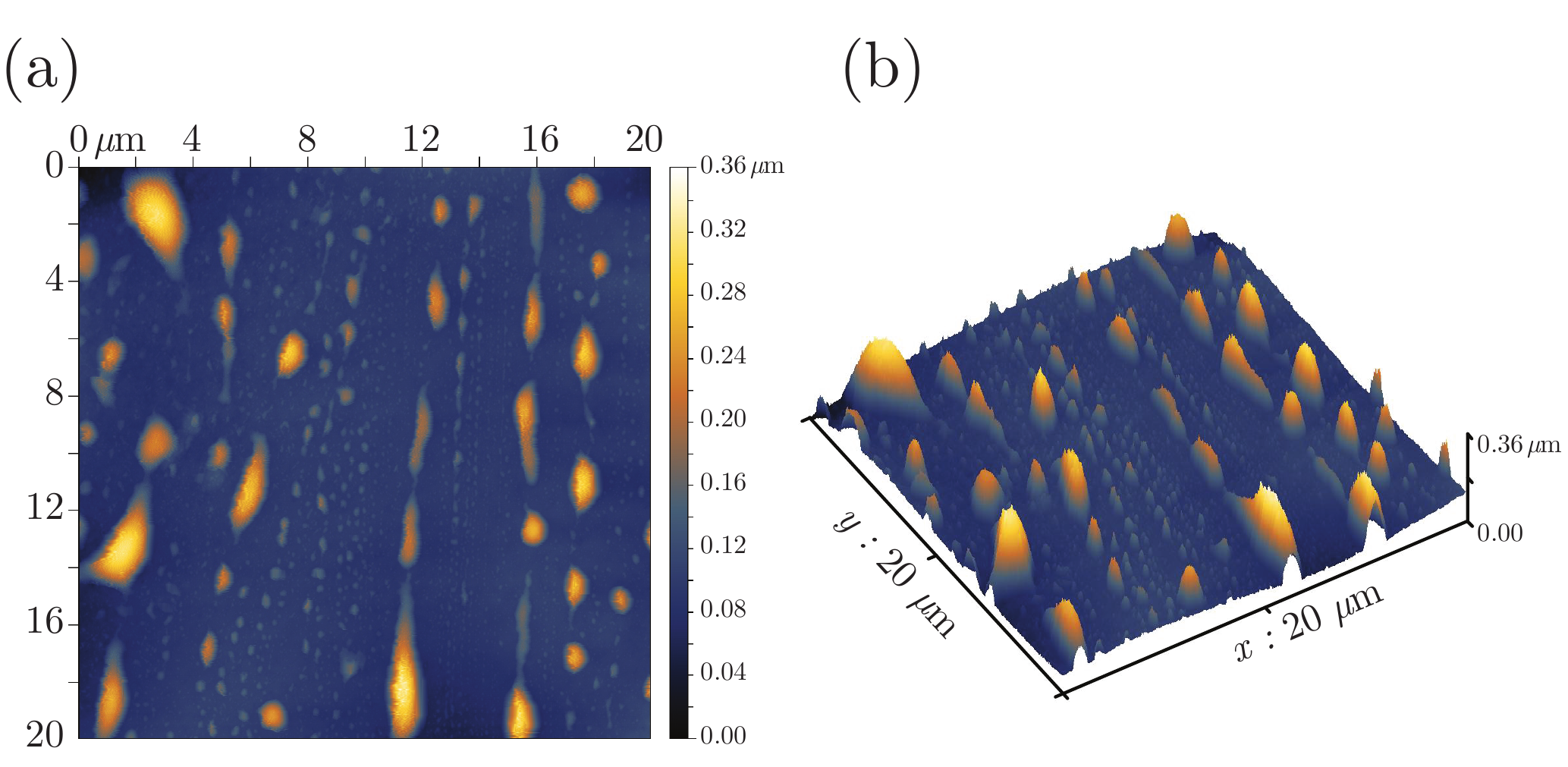}
\caption{\label{AFM_magnet} Topography of the magnet surface measured by Atomic Force Microscopy (a) in 2D and (b) 3D representation.}
\normalsize
\end{figure}

The difference between the Hertzian predicted shear stiffness, which is obtained considering smooth surfaces in contact, and the experimentally calculated one could be explained by the rough surface of the permanent magnets. By an Atomic Force Microscopy (AFM) roughness measurement, see Fig~\ref{AFM_magnet}, we found that the height of the asperities is around $0.2$-$0.3$~$\mu$m.
We also measured an averaged distance between the asperities around 2-3~$\mu$m. Thus, the contact between spheres and magnets can be considered as a multicontact interface. 
In this case, as it has been shown in \cite{shearrough}, the shear stiffness varies proportionally to the normal load $F$, $\tilde{K}_S=F/\lambda$ where $\lambda$ is an elastic length that lies in the micrometer range, and which is the relevant scale for the roughness of the surfaces. In our case, considering $\lambda=3~\mu$m (the average distance between the asperities) and $F=10$~N, we obtain $\tilde{K}_S=3.3\cdot 10^6$~N/m, very close to the experimentally obtained value. 
On the other hand, the roughness of the steel 440C beads is provided to be
0.025~$\mu$m much smaller than the static overlap predicted by the Hertzian theory using the experimentally evaluated values of the static loads $F$ and $F_0$. Thus, the contact between the beads can be considered as single contact.
 Another point that we should comment is
the fact that the magnets are fixed to one extremity (glued to the substrate) and free to move at the other extremity where the beads are located. The magnets can partially follow the bead motion, and as a consequence, the shear rigidity between bead and magnet is not completely activated. By FEM simulations we found that the first flexural motion of the cylindrical magnet is around $3.4$~kHz not exactly around $4.4$~kHz which is one of the gaps around the ZGV in Fig.~\ref{fig_Dis_exp_B2_Magnet_15beads}(a). Thus the gap cannot be caused by the hybridization of the chain modes with the vibrations of the magnets. To further test this assumption, we performed experiments using shorter cylindrical magnets of 4~mm height and the same magnetization. FEM simulations predict the first flexural mode to be at $22.7$~kHz, but once again the  $\tilde{K}_S$ is weaker than the Hertzian predicted value.
Thus, we believe that this disagreement in the $\tilde{K}_S$ is connected with the multicontact interface between the surface of the magnet and the sphere. This actually reveals another important feature in the proposed configuration: the tunability of the shear stiffness and thus of the dispersion relation by properly designed surface topographies. The latter can be produced either through a slow etching process using $H_2$ plasma \cite{microsilica} or by controlled abrasive powder coating \cite{shearrough}.

\section{Conclusion}
We have studied the  propagation properties of elastic waves in a magneto-granular phononic crystal consisting of a granular chain in contact with an array of fixed magnets.
Taking into account all degrees of freedom in transversal and rotational motions, we were able to derive the full dispersion relation. We have theoretically shown that our system supports transversal/rotational propagation modes, zero group velocity (ZGV) modes and accidental degeneracies in the dispersion curves.
Finally, we experimentally realized such a magneto-granular phononic crystal composed of spherical ferromagnetic spheres inside a properly designed magnetic field induced by  permanent magnets with which the beads are in contact. By fitting only one parameter, the shear stiffness between the spheres and the magnets, we obtained a very good agreement between experimental results and the predicted allowed and forbidden bands of propagation. We were also able to experimentally observe a ZGV point at finite wavelength. Disagreement between the shear stiffness predicted by the Hertzian theory of single contact and the experimentally obtained is attributed to the rough surface of the magnet, which was characterized using AFM techniques. 
The use of properly designed magnetic fields and particular surface topographies can lead to a programmable control of the stiffnesses between the elements of the setup and thus to an engineered dispersion relation. In addition, one can also modify the contact stiffnesses by placing dielectric spacer between the magnets and the spheres. This will increase the distance between the magnets and the spheres, resulting to weaker contact forces between the spheres and the dielectric spacer. 
 
Engineering the dispersion relation of the magneto-granular phononic crystal in combination with the nonlinear response of the Hertzian contact, could open the way for novel nonlinear and tunable mechanical metamaterials for the control of wave propagation.

\begin{acknowledgments}
G. T. acknowledges financial support from FP7-CIG (Project 618322 ComGranSol). We would like also to thank the team 3MPL of the IMMM for helping us with the AFM roughness surface characterization and S. Job for discussions.
\end{acknowledgments}

\appendix

\nocite{*}


\end{document}